\documentclass[aa,floatfix]{revtex4}
\usepackage{graphics,epsfig}
\usepackage{graphicx}
\usepackage{txfonts}

\begin{document}

\newcommand{\be}{\begin{equation}}
\newcommand{\ee}{\end{equation}}
\newcommand{\mf}{\mathbf}
\newcommand{\bee}{\begin{eqnarray}}
\newcommand{\een}{\end{eqnarray}}

\title{A Coarse-Grained  Field Theory for Density Fluctuations and
 Correlation Functions of Galactic Objects}
\author{Yang Zhang \cite{email}}
\affiliation{Astrophysics Center, University of Science and
Technology of China, Hefei, Anhui,  China}

\begin{abstract}
  {}
{To investigate the large scale structure of the universe from a
field-theoretical perspective, and to give a unifying
understanding of several seemingly-unrelated observed features of
clusterings of galactic objects, such as the dependence of
clustering amplitude on mass and luminosity, the departure of the
correlation function $\xi(r)$ from the power-law $\propto
(r_0/r)^{1.7}$, the oscillatory behavior of $\xi_{cc}(r)$ for
clusters on very large scales, and the scaling of $r_0$ with the
mean separation $d$ of clusters.
 }

  {We present a coarse-grained field theory of density fluctuations
for a Newtonian self-gravitating many-body system, and apply it to
a homogeneous Universe with small density fluctuations. The theory
uniformly treats the clustering of all types of galactic objects
in terms of the field of density fluctuations.}

{The Jeans length $\lambda_0$, a unique physical scale for a
gravitating system, appears naturally as the characteristic scale
underlying the large scale structure. Under Gaussian approximation
the analytic expressions of $\xi(r)$ and $P(k)$ are obtained. The
correlation amplitude is proportional to the galactic mass,
$\xi(r)\propto m$, and $\xi(r)$ departures from the power-law
$(r_0/r)^{1.7}$ around $r\sim 20$ $h^{-1}$ Mpc, and is oscillating
over large scales $\sim 100$ $h^{-1}$Mpc and damped to zero . The
spectrum amplitude is inversely proportional to the galactic
number density, $P(k)\propto 1/n$, and $P(k) \propto k^{ -2.2}$ in
the range $k \simeq ( 0.05\sim 0.3)$ hMpc$^{-1}$. The scaling $r_0
\propto d^{0.3\sim 0.5}$ holds quantitatively.}

{These preliminary analytic results under Gaussian approximation
already qualitatively explain these pronounced features of the
observed large scale structures. Yet on small scales the predicted
clustering is insufficient, and the peak of $P(k)$ is too sharp.
These deficiencies are due to neglecting higher order nonlinear
effects, which need to study further. }
   {}
\end{abstract}

\maketitle

 ~



\large

\baselineskip=13truept

\section{Introduction}

In past decades,
great progresses have been made in
understanding the the large scale structure of the universe.
Not only observations of the major galaxy surveys
such as  SDSS, 2dF, APM,  et. al.,
(Tegmark 2004; Zehavi 2005;
Colless 2001; Hawkins 2003; Madgwick 2003; Percival 2005;
Maddox 1996; Padilla 2003)
have revealed
the cosmic structures of increasingly large sky dimension,
theoretical studies have also achieved important results,
through both analytical modelling and numerical simulations.
The correlation functions of galactic objects,
are key physical quantities,
which contain important information, both
dynamic and statistical, of the cosmic structure.
Observations have shown that
the correlation functions of various galactic
objects, such as galaxies, groups of galaxies, quasars, and
clusters, are usually fitted by a single canonical power-law $
\xi(r) = (r_0/ r) ^{\gamma} $ in the range $r \lesssim 4 r_0$ with
the index $\gamma \sim 1.7$ and the  `correlation scale' $r_0 \sim
5 h^{-1}$Mpc for galaxies, $\sim 10 h^{-1}$Mpc for galaxy groups,
$\sim 20h^{-1}$Mpc for clusters, respectively
 (Peebles 1993; Bahcall 1996; Bahcall et al 2003).
Also $r_0$ increases with mass, richness, and luminosity of
galactic objects. Moreover, $r_0$ scales with the mean cluster
separation $d$ as $r_0(d) \propto d^{\,  0.3\sim 0.5}$
(Bahcall 1996; Croft 1997; Gonzalez 2002; Bahcall et al 2003;
 Zandivarez, Merchan, Padilla 2003).
On large scales $\xi(r)$ departures from the
power-law, drops steeply to zero, and seems to have an oscillating
behavior on very large scales $\geq 100h^{-1}$Mpc. These  observed
features have been  thought to be a theoretical challenge
(Bahcall 1996; Bahcall et al 2003).
From  view point of dynamics, the
Universe filled with various galactic objects is a many-body
system with attractive gravity between bodies.
A powerful,
systematic  approach to deal with statistical dynamics of this
kind of systems is through introducing the coarse-graining order
parameter, which converts the degrees of freedom of discrete
particles into the effective continuous field.
Therefore, a field theory arises, which combines
both the statistical and dynamical properties of the system.
In particular, the
method has been very successful in revealing the correlation
properties of fluctuations for a variety of homogeneous systems
near phase transition (Goldenfeld 1992).
Therefore, it should be
appropriate to employ the method to study the density fluctuations
of the homogeneous universe as a self-gravitating systems. In this
paper we will formulate such a field theory of density
fluctuations and apply it to the Universe on large scales. It
turns out that our analytic result at the level of Gaussian
approximation already interprets qualitatively the above-mentioned
several features  simultaneously.
Different from the conventional analytical studies,
such as the perturbation theory (Bernardeau 2001)
and the thermodynamic formulation (Saslaw 2000),
our approach is a field-theoretical study,
like those used in condensed matter and in particle physics.
For instance, in our theory,
the 2-point correlation function turns out to be the Green's function,
and the differential equation that it satisfies
can derived explicitly.

\section{  Field Equation of Correlation Function }

Consider a system  of $N$ galaxies of mass $m$ in the Universe
with the Hamiltonian
$H = \sum_{i}^{N} p_i^2/2m -   \sum_{i<j}^N Gm^2/|\mf{r_i-r_j}| $.
The grand partition function of this system at temperature $T$ is
\begin{equation}
Z =  \sum_{N=0}^{\infty}
\frac{z^N}{N!} \int \prod_{i=1}^N \frac{d^3p_i\,d^3r_i}{(2\pi)^3} e^{-H/T},
\end{equation}
where  $z$ is the fugacity.
By using the Hubbard-Stratonovich transformation
(Zinn-Justin  1996)
it can be cast into a path integral:
$Z = \int D\phi
\exp{[- \beta^{-1} \int d^3 {\mf{r}} {\cal L}(\phi)]}$,
where  $\phi$ is the gravitational field,  $ \beta \equiv   4\pi Gm/c_s^2$
of dimension $m^{-1}$, $ c_s \equiv (T/m)^{1/2} $
the sound speed,  the effective Lagrangian is
${\cal L}(\phi)=\frac{1}{2}({\bigtriangledown} \phi)^2 -k_J^2 e^{\phi}$,
and $  k_J \equiv  (4\pi G\rho_0/c_s^2)^{1/2}$ is the Jeans wavenumber.
The term $-k_J^2 e^\phi $  has a minus sign because gravity is attractive.
By the Poisson's equation
$\bigtriangledown ^2 \phi ({\mf{r}}) + k_J^2 e^{\phi}=0$
the mass density $\rho$ is related to the $\phi$ field by
$\rho ({\mf{r}})  = mn({\mf{r}}) = \rho_0   e^{\phi(\mf r)}$.
So  $ \rho_0$ is the constant mass  density when $\phi =  0$.
We define  a  dimensionless re-scaled mass density field
\begin{equation}
\psi ({\mf{r}}) \equiv  e^{\phi ({\mf{r}})}
 =  \rho ({\mf{r}})/\rho_0,
 \end{equation}
viewed as the coarse-grained order parameter,
and introduce a Schwinger type of  external source $J$
coupling with $\psi$ in the effective Lagrangian
\begin{equation}
{\cal L}(\psi, J)  = \frac{1}{2}
\left( \frac{\bigtriangledown \psi}{\psi} \right)^2 -k_J ^2  \psi -J\psi.
\end{equation}
The   source $J$ is used
to handle the functional derivatives with ease.
The  equation for  $\psi$  in the presence of  $J$  is
\be  \label{psij}
\bigtriangledown ^2 \psi -\frac{1}{\psi}(\bigtriangledown \psi)^2
+ k_J ^2 \psi^2 +J\psi^2 = 0 ,
\ee
the generating functional is
\begin{equation}
Z[J] = \int D\psi \exp{[-\beta^{-1}\int d^3{\mf{r}} {\cal L}(\psi, J)] },
\end{equation}
the $n$-point connected  correlation function is
\begin{eqnarray}
 G^{(n)}_c ({\mf{r}}_1, \ldots , {\mf{r}}_n)
= \langle \delta\psi ({\mf{r}}_1 )\ldots  \delta\psi({\mf{r}}_n)\rangle
  \nonumber  \\
=  \beta^n \frac{\delta ^n \ln Z[J]}{ \delta J({\mf{r}}_1)...
\delta J({\mf{r}}_n)  }|_{J=0},
\end{eqnarray}
where $ \delta\psi ({\mf{r}})= \psi( {\mf{r}})-\langle\psi(\mf r) \rangle$
is the fluctuation field about the expectation value
$\langle\psi(\mf r) \rangle$.
The standard way to obtain the equation
for the 2-point correlation function
$G^{(2)} _c({\mf{r}})$ is the following
(Goldenfeld 1992).
Taking expectation value of Eq.(\ref{psij}) in the presence
of $J$, using the approximation
$\langle (\bigtriangledown \psi)^2/\psi  \rangle _J \simeq
(\bigtriangledown \langle\psi \rangle _J)^2/ \langle\psi\rangle_J$,
which is equivalent to neglecting high order terms of $\delta\psi$,
taking functional derivative of this equation  w.r.t. the
source $J$, and setting  $J=0$, we end up with
\begin{eqnarray}
 \label{ge}
\bigtriangledown ^2  G^{(2)} _c({\mf{r}})  + & &
(\frac{\bigtriangledown \langle\psi \rangle}{\langle\psi \rangle} )^2
                      G^{(2)}_c({\mf{r}})
      -\frac{2}{\langle\psi \rangle} \bigtriangledown
\langle\psi \rangle   \cdot \bigtriangledown G^{(2)} _c({\mf{r}})
   \nonumber  \\
& & +      2 k_J ^2 \langle\psi \rangle G^{(2)} _c({\mf{r}}) =
-  \langle\psi \rangle^2   \beta \delta^{(3)}({\mf{r}})  ,
\end{eqnarray}
where $ G^{(2)}_c (  {\mf{r- r'}}) =
\beta \delta   \langle \psi({\mf{r}})\rangle_{J=0} / \delta J({\mf{r'}})$
has been used,
and $\langle\psi \rangle \equiv \langle\psi \rangle_{J=0}$ is the
background field depending on the position $\mf{r}$ in general.
We remark that when higher order terms in
$1/\psi\simeq 1/\psi_0+\delta\psi/\psi_0^2+...$
were kept,
one would end up with nonlinear equation
much more complected than Eq.(\ref{ge}).
Assume now  the homogeneity of the Universe on large scales.
Then the background density field is constant
$ \langle \psi  \rangle = \psi_0 = 1 $.
Then  Eq.(\ref{ge}) reduces to the Helmholtz equation with a point source:
\begin{eqnarray}  \label{eq-g}
\bigtriangledown ^2  \xi({\mf{r}})
+2 k_J ^2  \xi({\mf{r}})
 = - \frac{4\pi Gm}{c_s^2} \delta^{(3)}({\mf{r}})  ,
\end{eqnarray}
where we have used a simple notation $\xi({\mf{r}})\equiv G^{(2)}_c(\mf r)$.
The approximation level of Eq.(\ref{eq-g}) is essentially
that of the Gaussian model in the Landau-Ginzburg theory(Goldenfeld 1992).
The  solution is the Green's function
\be \label{g}
\xi(r) = A \left(  c \frac{\cos (k_0r)}{ r} +
(1-c)\frac{\sin (k_0r)}{ r} \right),
\ee
where the amplitude $A\equiv  Gm/c_s^2$,
the  parameter  $c$
determines the  contributions from  independent $cos$  and $sin$ modes,
and the characteristic wavenumber $k_0 \equiv   \sqrt{2}k_J$.
Thus the wavelength $ \lambda_0  = \lambda_J/ \sqrt{2}$
occurs as the characteristic scale that underlies the large scale structure,
where $ \lambda_J $ is the Jeans length.
The `correlation scale' $r_0$ in the canonical power-law
is only empiric not fundamental.
The amplitude $A \propto  G m/c_s^2  r_0$,
the ratio of the gravitational correlation
energy $Gm^2/r_0$ to twice  the kinetic energy $mc_s^2$ of
peculiar motion of galaxies.
So a greater  gravitational energy yields
a larger correlation amplitude, similar to the conclusion
by Saslaw (Saslaw 2000).
A comparison with the Landau-Ginzburg theory will be suggestive.
Recall that for a fluid,
with the dominant force between molecules being  repulsive,
the Gaussian approximation yields
a correlation function $\propto e^{-r/\eta}/r$,
where  $\eta$ is the correlation length of
phase transition (Goldenfeld 1992).
Contrarily, Eq.(\ref{g}) is oscillating, instead of decaying,
because gravity is attractive.
In this sense, the self-gravitating system is always
at  the critical point of `phase transition'
with $\eta \rightarrow \infty$ (Saslaw 2000).
Therefore, it is pertinent to study the  density fluctuations
of  the homogeneous universe
within the framework of the field theory,
where  the coarse-grained order parameter
is the rescaled density field $\psi$.
If the gravity were repulsive,
the term $2 k^2_J\psi$ in Eq.(\ref{eq-g}) would have a minus sign,
and Eq.(\ref{g}) would be of
the same form $\xi(r) \propto e^{-|k_0|r}/r$ as for a fluid.

Using the definition of the projected correlation function
(Davis, Peebles 1983)
$
w_p(r_p)=2\int^{\infty}_{r_p}
\xi (r)(r^2-r_p^2)^{-1/2}\, r dr,
$
and Eq.(\ref{g}),
one gets
$
\label{projcorr}
w_p(r_p)   = \pi A[  -c N_0(k_0  r_p)  + (1-c) J_0(k_0 r_p)  ],
$
where $N_0$ is the Neumann function and $J_0$ the Bessel function.
So the amplitude is also proportional to the mass,
$w_p \propto m$,
and $w_p(r_p)$ has a strong departure from the power-law
and drops steeply  to zero, agreeing with the observations
(Budavari 2003, Zehavi et al 2002; 2005).

\section{ Implications}

Now examine some implications of $\xi (r)$ for the large scale structure.
Although, as the Gaussian approximation,
Eq.(\ref{g}) does  not yet give a detailed account of
of large scale structure,
they already reveal the following pronounced features.

1) Galactic objects in our theory
are all treated as  gravitating particles,
and have the same form of correlation functions as Eq.(\ref{g}),
differing only in $A$ and $k_0$.
Indeed,  the observed $\xi (r)$ for galaxies, groups, and clusters
all have a similar shape,
for instance, the correlation function of RELEX clusters
divided by $\sim 10$
coincide with that of EPS galaxies (Guzzo 2000).

2)  Since the amplitude $ A \propto m$,
galactic objects with a greater mass have a higher correlation.
Besides, since richness is proportional to $ m$ by definition and
more luminous galaxies generally have greater mass,
this result naturally explains why the  correlation amplitudes
increase with richness and luminosity of galactic objects
without recourse to the bias model.
Attempts have been made to explain this in terms of the
statistics of rare peak events (Kaiser 1984).
Moreover, early type galaxies (ellipticals) have a mean number density $n$
lower than that of late type (spirals)
and are more likely located in denser regions,
thus they represent a greater mass of the region
(Davis, Geller 1976; Madgwick 2003)
and have a higher correlation.

3) $\xi(r)$ has a departure from the canonical power law,
as is observed.
Plot  $\xi (r)$ in FIG.\ref{graph1}
to fit with $\xi_{gg}(r)$
of the 2df galaxies (Madgwick 2003).
The parameters  $A \simeq 3$ h$^{-1}$Mpc, $c\simeq  1.7$ ,
and $k_0\simeq 0.05$ hMpc$^{-1}$ are taken.
It is seen that the  departure
from the power-law occurs at $\sim 8$ h$^{-1}$Mpc,
and $\xi (r)$ goes to zero at $ \simeq (20\sim 30)$ h$^{-1}$Mpc.
But on small scales $< 2$ h$^{-1}$Mpc,
$\xi (r)\propto r^{-1}$, lower than the observed.
This small-scale deficiency of correlation
is expected since significant clustering effects
of higher order terms of $\delta \psi$ have
been neglected in Eq.(\ref{eq-g}).
We expect these nonlinear effects will improve
this small-scale correlation.
For groups and clusters, fitting can be done similarly,
requiring a larger $A$ and a slightly smaller $k_0$.
 However, the redshift-space correlation function
from the SDSS luminous red galaxies crosses to zero
only after $100\, h^{-1}$Mpc (Eisenstein et al 2005; Huetsi 2006).
It is difficult for our theory,
and it might indicate that either the universe is still
less relaxed,
or the higher order nonlinear effects need to be taken into account.

4)  The shape of observed correlation function of
galaxies depends on their `color'.
The red galaxies are more luminous with a
slightly higher mean number density $n$,
and exhibit a steeper and higher-amplitude correlation function
than the blue galaxies
(Zehavi 2002; 2005).
This is explained by Eq.(\ref{g}),
which, for a larger $k_0 \propto (nm)^{1/2}$,
is steeper and has a higher amplitude on small scales.

5)   As mentioned at the beginning, the observations show that
the  `correlation scale' $r_0$ increases with the mean separation $d$
in such a way as
$r_0(d) \propto d ^{0.3\sim 0.5}$
(Bahcall 1996; Bahcall et al 2003; Croft 1997).
The mechanism underlying
was thought to be a scaling behavior
(Szalay, Schramm 1985).

In our theory,
if we adopt the power-law fitting $r_0(d) \propto\xi^{1/1.7}$ with
$\xi$ being given in (\ref{g}), then, for typical clusters,
$r_0(d) \sim 20 \cos^{1/1.7} \,[ (30/d)^{3/2} \,]$,
increasing in the range $d =10 \sim 80 $h$^{-1}$Mpc,
as shown in FIG.\ref{graph2}.
Although it is only a approximate result,
the trend of the profile is in qualitative agreement with the observation.

6) On  very large scales,
$\xi (r)$ is a damped oscillation about zero.
Some surveys have indicated this feature
with  $ \lambda_0   \simeq 120 \sim 200$ h$^{-1}$Mpc
(Broadhurst et al 1990; Tucker 1997;
Einasto et al 1997; 1997b; 2002;  2002b; Tago 2002;
Eisenstein et al 2005;
 Yanata 2005).
Plotting $\xi(r)$ in FIG.\ref{graph3} to fit to
the observed $\xi_{cc}(r)$ of clusters
requires $A\simeq 50$ h$^{-1}$Mpc and $k_0\simeq 0.05$ h Mpc$^{-1}$.
Thus the Gaussian approximation in Eq.(\ref{g}) gives a good description
of the oscillatory correlation on very large scales.

\section{ Power Spectrum}

The power spectrum is the Fourier transform of the correlation function,
and can be also derived directly by Fourier transforming  Eq.(\ref{eq-g}),
\be \label{power}
P(k) = \frac{1}{2n}\frac{1}{(\frac{k}{k_0})^2-1}.
\ee
Firstly, this predicts that
$ P(k) $ has a higher amplitude for galactic objects
with a smaller spatial number density $n$,
agreeing with a whole variety of surveys.
For instance, the observed $P(k)$ for all Abell clusters is lower
than that for Abell clusters with measured redshifts,
and the same fact is also found for APM clusters
(Einasto 2002b).
This is easily understood by Eq.(\ref{power})
as all clusters have a higher $n$ than those with measured redshifts.
Secondly, $P(k)$ has a sharp peak at $k=k_0$,
corresponding to the periodic wavelength $\lambda_0 =2\pi/k_0$.
Taking the observed value
$ \lambda_0  \simeq  120\sim 200$ h$^{-1}$Mpc yields
the peak location $k_0 \simeq  0.031 \sim 0.052$ hMpc$^{-1}$.
The peak has been, to some degree, indicated by several cluster surveys
(Ratcliffe et al 1996;
Tadros, Efstathiou, Dalton, 1998; Hoyle et al 1999;
Collins 2000;  Schuecker et al 2001; Carretti et al 2002;
 Einasto et al 2002; 2002b;
 Tago et al 2002; Padilla, Baugh 2003;
Tegmark 2004;  Percival 2005).

Thirdly, as shown in FIG.\ref{graph4},
for large scales $k \simeq ( 0.05\sim 0.3)$ hMpc$^{-1}$,
(\ref{power}) gives $P(k) \propto k^{ -2.2}$,
agreeing well with the major surveys
(Peacock 1997; 1999; Carretti et al 2002;  Dodelson et al 2002),
whereas   the power-law with $\gamma\sim 1.7$
would yield  $\propto  k^{-1.3}$.
But for small scales $ k \simeq  (0.4 \sim 3)$ hMpc$^{-1}$,
Eq.(\ref{power}) gives $P(k)\propto  k^{-2}$,
the observed is $ \propto k^{ -1.2}$
(Peacock 1999; Dodelson et al 2003).
This  small-scale deficiency is the same phenomenon
occurring in $\xi(r)$ noticed before,
also due to neglect of the nonlinear terms.
But there are two more difficulties for $P(k)$ in Eq.(\ref{power}).
It takes on negative values on large scales with $k < k_0$,
and its peak is too sharp.
We expect that,
when the nonlinear and frictional effects are properly taken into account,
the negative power at  $k<k_0$ may also be alleviated
and the peak may be smoothed to certain extent.
These  are obviously different from the observed power spectrum.
A real spectrum can not have an infinitely
sharp peak, which should be smoothed by some damping effects.
Recent observations have indicated that the measured $P(k)$
has a rather flat peak with
 large systematic errors on large scales
(Einasto et al 2002b;  Outram et al 2003;
Padilla, Baugh 2003;  Tegmark et al 2004;
Maller et al 2005; Percival 2005).

\section{ Conclusion }

The effective field theory for density fluctuations
presented in this paper has the Jeans scale
$k_0\simeq (0.04\sim 0.06)$ hMpc$^{-1}$  as the unique scale
underlying the large scale structure.
Even at the level of Gaussian approximation
its preliminary results already interpret
simultaneously several important
features of the observed large scale structures,
some of them have been thought to be puzzling and seemingly-unrelated.
But the Gaussian approximation
does not yet give a sufficient clustering on small scales.
Therefore, it is
worth while to explore further the  field theory
beyond the Gaussian approximation,
and to take into account of higher order terms of $\delta \psi$
neglected in Eq.(\ref{eq-g}).
This is expected to enhance the low amplitude
of clustering on small scales
and smooth out the sharpness of peaks in the power spectrum.
The study of these nonlinear effects is still underway.

This work was supported by the CNSF No.10173008,
 NKBRSF G19990754, and by SRFDP.

\begin{figure}
\caption{\label{graph1}
The calculated $\xi (r)$ in Gaussian approximation
compared with the observed  $\xi_{gg}(r)$
of 2dFGRS (Madgwick et al 2003).}
\end{figure}

\begin{figure}
\caption{\label{graph2}
The calculated scaling $r_0(d)$
compared with
the observation (Bahcall 1996; Bahcall et al 2003)}
\end{figure}

\begin{figure}
\caption{\label{graph3}
The large scale oscillations of
$\xi(r)$ compared with the observed $\xi_{cc}(r)$ of clusters
(Einasto et al 2002;  2002b).}
\end{figure}

\begin{figure}
\caption{\label{graph4} The analytic  $P(k)$ in Eq.(\ref{power})
compared with the SDSS galaxies (Dodelson et al 2002)
in the range $k\simeq (0.1 \sim 5)$hMpc$^{-1}$.  }
\end{figure}

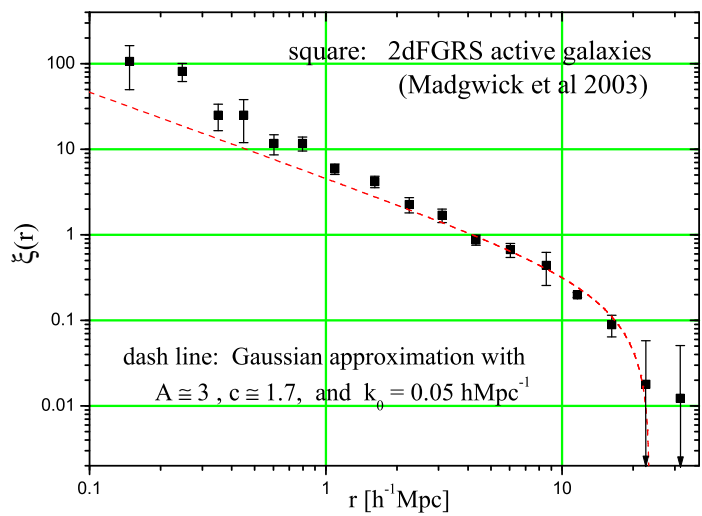
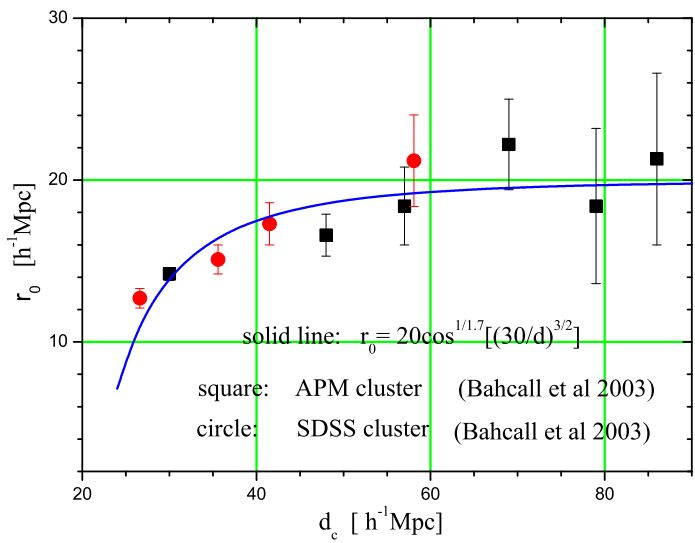
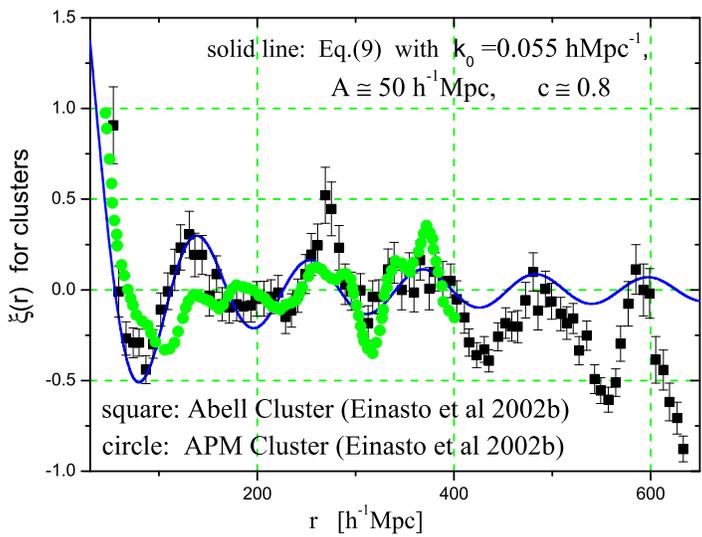
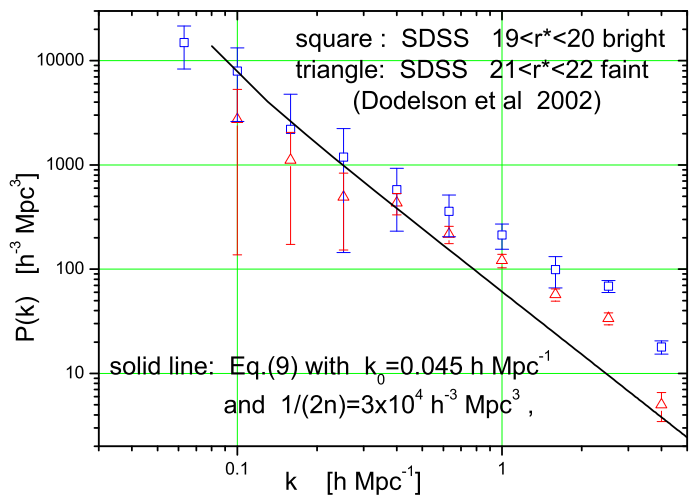
\end{document}